\documentclass[11pt]{article}
\usepackage{moriond,epsfig}

\def\D0bar{\overline D{}^0}
\def\K0bar{\overline K{}^0}
\def\etacp{\eta_{CP}}
\def\3bar{\overline{3}}
\def\sixbar{\overline{6}}
\def\tenbar{\overline{10}}
\def\15bar{\overline{15}}
\def\24bar{\overline{24}}
\def\42bar{\overline{42}}
\def\60bar{\overline{60}}
\def\cO{{\cal O}}
\newcommand{\BR}{{\cal B}}
\newcommand{\beq}{\begin{equation}}
\newcommand{\eeq}{\end{equation}}
\newcommand{\beqa}{\begin{eqnarray}}
\newcommand{\eeqa}{\end{eqnarray}}
\def\lesssim{\mathrel{\rlap{\lower4pt\hbox{\hskip.1pt$\sim$}}
  \raise1pt\hbox{$<$}}}		%less than or approx. symbol
\def\gtrsim{\mathrel{\rlap{\lower4pt\hbox{\hskip.1pt$\sim$}}
  \raise1pt\hbox{$>$}}}		%greater than or approx. symbol

\begin{document}
\vspace*{4cm}
\title{\boldmath $D^0 - \D0bar$\, MIXING}

\author{Zoltan Ligeti}

\address{Ernest Orlando Lawrence Berkeley National Laboratory\\
  University of California, Berkeley, CA 94720}

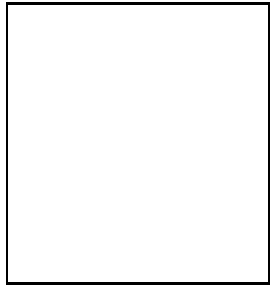
\maketitle\abstracts{
The main challenge in the Standard Model calculation of the mass and width
difference in $D^0-\D0bar$ mixing is to estimate the size of $SU(3)$ breaking. 
We prove that mixing occurs in the Standard Model only at second order in
$SU(3)$ violation.  We calculate $SU(3)$ breaking due to phase space effects,
and find that it can naturally give rise to a width difference $\Delta\Gamma /
2\Gamma \sim 1\%$, potentially reducing the sensitivity of $D$ mixing to new
physics.\,\cite{FGLP,BGLNP} \hfill LBNL-50404}

\section{Introduction}

It is a common assertion that the Standard Model (SM) prediction for mixing in
the $D^0-\D0bar$ system is very small, making this process a sensitive probe of
new physics.  Two physical parameters that characterize $D^0-\D0bar$ mixing are
\beq
x \equiv {\Delta M \over \Gamma}\,, \qquad
  y \equiv {\Delta \Gamma \over 2\Gamma}\,,
\eeq
where $\Delta M$ and $\Delta \Gamma$ are the mass and width differences of the
neutral $D$ meson mass eigenstates, and $\Gamma$ is their average width.
The $D^0-\D0bar$ system is unique among the neutral mesons in that it is the
only one whose mixing proceeds via intermediate states with down-type quarks,
and is therefore very sensitive to certain classes of new physics
models.\,\cite{sqa}\, The mixing is very slow in the SM, because the third
generation plays a negligible role due to the smallness of $|V_{ub} V_{cb}|$
and $m_b \ll m_W$, and so the GIM cancellation is very
effective.\,\cite{Ge92,Oh93,Da85,dght}

The current experimental upper bounds on $x$ and $y$ are of order $10^{-2}$,
and are expected to improve significantly in the coming years.  To regard a
future discovery of nonzero $x$ or $y$ as a signal for new physics, we would
need high confidence that the SM predictions lie significantly below the
present limits.  As we will show, $x$ and $y$ are generated only at second
order in $SU(3)$ breaking in the SM, so schematically
\beq
x\,,\, y \sim \sin^2\theta_C \times [SU(3) \mbox{ breaking}]^2\,,
\eeq
where $\theta_C$ is the Cabibbo angle.  Therefore, the SM values of
$x$ and $y$ depend crucially on the size of $SU(3)$ breaking. 
Although $y$ is expected to be determined by SM processes, its value affects
significantly the sensitivity of $D$ mixing to new physics.\,\cite{BGLNP}

At present, there are three types of experiments which measure $x$ and $y$. 
Each is actually sensitive to a combination of $x$ and $y$, rather than to
either quantity directly.  First, the $D^0$ lifetime difference to $CP$ even
and $CP$ odd final states\,\cite{e791y,focusy,cleoy,belley,babary} measures,
to leading order,
\beq\label{ycp}
y_{CP} = {\Gamma(CP \mbox{ even}) - \Gamma(CP \mbox{ odd}) \over
  \Gamma(CP \mbox{ even}) + \Gamma(CP \mbox{ odd})}
  \simeq {\hat \tau(D \to \pi^+ K^-) \over \hat\tau(D \to K^+ K^-)} - 1
  = y\cos\phi - x \sin\phi\, \frac{A_m}2\,,
\eeq
where the $D$ mass eigenstates are $|D_{L,S}\rangle = p\, |D^0\rangle \pm q\,
|\D0bar\rangle$, $A_m = |q/p|^2-1$, and $\phi = \mbox{arg}(q/p)$ is a possible
$CP$ violating phase of the mixing amplitude.  The experimental results are
\begin{equation}
\quad y_{CP} = \cases{ 
  \phantom{-}0.8 \pm 3.1 \%  &  E791\,,\,\cite{e791y} \cr
  \phantom{-}3.4 \pm 1.6 \%  &  FOCUS\,,\,\cite{focusy} \cr
            -1.1 \pm 2.9 \%  &  CLEO\,,\,\cite{cleoy} \cr
            -0.5 \pm 1.3 \%  &  BELLE\,,\,\cite{belley} \cr
  \phantom{-}1.4 \pm 1.2 \% \qquad  &  BABAR\,,\,\cite{babary} \cr}
\end{equation}
which yield a world average $y_{CP} = 1.0 \pm 0.7 \%$ at present.  Second, the
time dependence of doubly Cabibbo suppressed decays, such as $D^0\to K^+
\pi^-$,\,\cite{Godang:2000yd} is sensitive to the three quantities
\beq\label{xyprime}
(x \cos\delta + y \sin\delta) \cos\phi\,, \qquad
  (y \cos\delta - x \sin\delta) \sin\phi\,, \qquad
  x^2 + y^2\,,
\eeq
where $\delta$ is the strong phase between the Cabibbo allowed and doubly
Cabibbo suppressed amplitudes.  A similar study for $D^0\to K^-\pi^+\pi^0$
would allow the strong phase difference to be extracted simultaneously from the
Dalitz plot analysis.\,\cite{brand}\, Third, one can search for $D$ mixing in
semileptonic decays,\,\cite{DmixSL} which is sensitive to $x^2+y^2$.

In a large class of models, the best hope to discover new physics in $D$ mixing
is to observe the $CP$ violating phase, $\phi_{12} = \mbox{arg}
(M_{12}/\Gamma_{12})$, which is very small in the Standard Model.  However, if
$y \gg x$, then the sensitivity of any physical observable to $\phi_{12}$ is
suppressed, since $A_m \propto x/y$ and $\phi \propto (x/y)^2$, even if new
physics makes a large contribution to $M_{12}$ and
$\phi_{12}$.\,\cite{BGLNP}\,  It is also clear from Eq.~(\ref{xyprime}) that
if $y$ is significantly larger than $x$, then $\delta$ must be known very
precisely for experiments to be sensitive to new physics in the terms linear in
$x$ and $y$.

\section{$SU(3)$ analysis of $D^0-\D0bar$ mixing}

In this section we prove that $D^0-\D0bar$ mixing arises only at second order
in $SU(3)$ breaking effects.  The proof is valid when $SU(3)$ violation enters
perturbatively.  The quantities $M_{12}$ and $\Gamma_{12}$, which determine $x$
and $y$, depend on matrix elements of the form $\langle\D0bar|\, {\cal H}_w
{\cal H}_w\, |D^0\rangle$, where ${\cal H}_w$ is the $\Delta C=-1$ part of the
weak Hamiltonian.  Denoting by $D$ the field operator that creates a $D^0$
meson and annihilates a $\D0bar$, this matrix element may be written as
\beq\label{melm}
  \langle 0| D\, {\cal H}_w {\cal H}_w D\, |0 \rangle\,.
\eeq
Let us focus on the $SU(3)$ flavor group theory properties of this expression.

Since the operator $D$ is of the form $\bar cu$, it transforms in the
fundamental representation of $SU(3)$, which we represent with a lower index,
$D_i$ (the index $i = 1,2,3$ corresponds to $u,d,s$, respectively).  The only
nonzero element is $D_1=1$.  The $\Delta C=-1$ part of the weak Hamiltonian has
the flavor structure $(\bar q_ic)(\bar q_jq_k)$, so its matrix representation
is written with a fundamental index and two antifundamentals, $H^{ij}_k$.  This
operator is a sum of irreducible representations contained in the product $3
\times \3bar \times \3bar = \15bar + 6 + \3bar + \3bar$.  In the limit in which
the third generation is neglected, $H^{ij}_k$ is traceless, so only the
$\15bar$ (symmetric in $i$ and $j$) and 6 (antisymmetric in $i$ and $j$)
appear.  That is, the $\Delta C=-1$ part of ${\cal H}_w$ may be decomposed as
${1\over2} (\cO_{\15bar} + \cO_6)$, where\,\cite{FGLP}
\beqa
\cO_{\15bar} &=& (\bar sc)(\bar ud) + (\bar uc)(\bar sd)
  + s_1(\bar dc)(\bar ud) + s_1(\bar uc)(\bar dd)\nonumber\\
&&{} - s_1(\bar sc)(\bar us) - s_1(\bar uc)(\bar ss)
  - s_1^2(\bar dc)(\bar us) - s_1^2(\bar uc)(\bar ds) \,, \nonumber\\
\cO_6 &=& (\bar sc)(\bar ud) - (\bar uc)(\bar sd)
  + s_1(\bar dc)(\bar ud) - s_1(\bar uc)(\bar dd)\nonumber\\
&&{} - s_1(\bar sc)(\bar us) + s_1(\bar uc)(\bar ss)
  - s_1^2(\bar dc)(\bar us) + s_1^2(\bar uc)(\bar ds) \,,
\eeqa
and $s_1=\sin\theta_C\approx0.22$.  This determines the nonzero elements of 
the matrix representations of $H(\15bar)^{ij}_k$ and $H(6)^{ij}_k$.
We introduce $SU(3)$ breaking through the quark mass operator ${\cal M}$, whose
matrix representation is $M^i_j={\rm diag}(m_u,m_d,m_s)$. Although ${\cal M}$
is a linear combination of the adjoint and singlet representations, only the 8
induces $SU(3)$ violating effects.  It is convenient to set $m_u=m_d=0$ and let
$m_s\ne0$ be the only $SU(3)$ violating parameter.  All nonzero matrix elements
built out of $D_i$, $H^{ij}_k$ and $M^i_j$ must be $SU(3)$ singlets.

We are now ready to prove that $D^0-\D0bar$ mixing arises only at second order
in $SU(3)$ violation, by which we mean second order in $m_s$.  First, note that
the pair of $D$ operators is symmetric, and so $D_i D_j$ transforms as a 6
under $SU(3)$.  Second, the pair of ${\cal H}_w$'s is also symmetric, and the
product $H^{ij}_kH^{lm}_n$ is in one of the representations which appears in
the product
\beqa
\left[ (\15bar+6)\!\times\!(\15bar+6) \right]_S &=&
  (\15bar\times\15bar)_S +(\15bar\times 6)+(6\times 6)_S \\*
&=& (\60bar+\24bar+15+15'+\sixbar) + (42+24+15+\sixbar+3)
  + (15'+\sixbar)\,. \nonumber
\eeqa
A straightforward computation shows that only the $\60bar$, 42, and $15'$
representations appear in the decomposition of ${\cal H}_w{\cal H}_w$.  So we
have product operators of the form
\beq
DD = {\cal D}_6\,, \qquad
  {\cal H}_w {\cal H}_w = \cO_{\60bar}+\cO_{42}+\cO_{15'}\,,
\eeq
where the subscripts denote the representation of $SU(3)$.

Since there is no $\sixbar$ in the decomposition of ${\cal H}_w{\cal H}_w$,
there is no $SU(3)$ singlet which can be made with ${\cal D}_6$, and so there
is no $SU(3)$ invariant matrix element of the form in Eq.~(\ref{melm}).  This
is the well-known result that $D^0-\D0bar$ mixing is prohibited by $SU(3)$
symmetry.

Now consider a single insertion of the $SU(3)$ violating spurion ${\cal M}$.
The combination ${\cal D}_6{\cal M}$ transforms as $6\times 8 = 24 + \15bar + 6
+ \3bar$.  There is still no invariant to be made with ${\cal H}_w {\cal
H}_w$.  This proves that $D^0-\D0bar$ mixing is not induced at first order in
$SU(3)$ breaking.

With two insertions of ${\cal M}$, it becomes possible to make an $SU(3)$
invariant.  The decomposition of ${\cal D}{\cal M}{\cal M}$ is
\beqa
6\times(8\times 8)_S &=& 6\times(27+8+1)\nonumber\\
  &=& (60+\42bar+24+\15bar+\15bar'+6) + (24+\15bar+6+\3bar) + 6\,.
\eeqa
There are three elements of the $6\times 27$ part which can give invariants
with ${\cal H}_w{\cal H}_w$.  Each invariant yields a contribution
proportional to $s_1^2m_s^2$.  As promised, $D^0-\D0bar$ mixing arises only at
second order in the $SU(3)$ violating parameter $m_s$.

\section{Estimating the size of $SU(3)$ breaking}

There is a vast literature on estimating $x$ and $y$ within and beyond the
Standard Model, and the results span many orders of
magnitudes.\,\cite{nelson}\,  Roughly speaking, there are two approaches,
neither of which is very reliable, because $m_c$ is in some sense intermediate
between heavy and light.

{\bf ``Inclusive" approach}~~ 
The inclusive approach is based on the operator product expansion (OPE).  In
the $m_c \gg \Lambda$ limit, where $\Lambda$ is a scale characteristic of the
strong interactions, such as $m_\rho$ or $4\pi f_\pi$, $\Delta M$ and
$\Delta\Gamma$ can be expanded in terms of matrix elements of local
operators.\,\cite{Ge92,Oh93,Bigi:2000wn}\,  The use of the OPE relies on local
quark-hadron duality, and on $\Lambda/m_c$ being small enough to allow a
truncation of the series after the first few terms.  However, the charm mass
may not be large enough for these to be good approximations for nonleptonic $D$
decays.  While an observation of $y$ of order $10^{-2}$ could be ascribed to a
breakdown of the OPE or of duality,\,\cite{Bigi:2000wn} such a large value of
$y$ is not a generic prediction of OPE analysis. 

The leading contribution in the OPE comes from 4-quark operators in the
$|\Delta C|=2$ effective Hamiltonian, corresponding to the short distance box
diagram.  The result is of the form
\beq\label{msd}
\Delta M_{\rm box} \propto {2X_D \over 3\pi^2}\, {m_s^4 \over m_c^2}\,, \qquad
  \Delta \Gamma_{\rm box} \propto {4X_D \over 3\pi}\, {m_s^6 \over m_c^4}\,,
\eeq
where $X_D = |V_{cs} V_{cd}^*|^2 G_F^2\, m_D f_D^2 B_D$. 
Eq.~(\ref{msd}) then yields the estimates
\beq
x_{\rm box}\sim {\rm few}\times 10^{-5}\,,\qquad
  y_{\rm box}\sim {\rm few}\times 10^{-7}\,.
\eeq
The $m_s^6$ dependence of $\Delta \Gamma_{\rm box}$ comes from three sources: 
(i) $m_s^2$ from an $SU(3)$ violating mass insertion on each quark line in the
box graph; (ii) $m_s^2$ from an additional mass insertion on each line to
compensate the chirality flip from the first insertion; (iii) $m_s^2$ to lift
the helicity suppression of the decay of a scalar meson into a massless fermion
pair.  The last factor of $m_s^2$ is absent from $\Delta M$; this is why at
leading order in the OPE, $y_{\rm box} \ll x_{\rm box}$.

\begin{table}
\caption{The enhancement of $\Delta M$ and $\Delta\Gamma$ relative to the box
diagram (i.e., the 4-quark operator) contribution at higher orders in the OPE. 
$\Lambda$ denotes a typical hadronic scale around 1\,GeV, and $\beta_0 =
11-2n_f/3 = 9$.}
\label{opetable}
\vspace{4pt}
\begin{center}
\begin{tabular}{|c|ccc|} \hline
ratio  &  4-quark  &  6-quark  &  8-quark \\ \hline
~~$\Delta M/\Delta M_{\rm box}$~~  &  1  &  ~~$\Lambda^2 /m_s m_c$~~
  &  ~~$(\Lambda^2/m_s m_c)^2({\alpha_s/4\pi})$~~ \\ \hline
$\Delta \Gamma/\Delta M$  &  ~~~~$m_s^2/m_c^2$~~~~
  &  $\alpha_s/4\pi$  &  $\beta_0\,\alpha_s/4\pi $ \\ \hline
\end{tabular}
\end{center}
\end{table}

Higher order terms in the OPE are important, because the chiral suppressions
can be lifted by quark condensates instead of mass insertions,\,\cite{Ge92}
allowing $\Delta M$ and $\Delta\Gamma$ to be suppressed by $m_s^2$ only.  This
is the minimal suppression required by $SU(3)$ symmetry.  The order of
magnitudes of the resulting contributions are summarized in
Table~\ref{opetable}.  The dominant contributions to $x$ are from 6- and
8-quark operators, while the dominant contribution to $y$ is from 8-quark
operators.  With some assumptions about the hadronic matrix elements, one finds
\beq 
x \sim y \sim 10^{-3}\,. 
\eeq 
A generic feature of OPE based analyses is that $x\gtrsim y$.  We emphasize
that at the present time these methods are useful for understanding the order
of magnitude of $x$ and $y$, but not for obtaining reliable results.  Turning
these estimates into a systematic computation of $x$ and $y$ would require the
calculation of many nonperturbative matrix elements.

{\bf ``Exclusive" approach}~~
A long distance analysis of $D$ mixing is complementary to the OPE.  Instead of
assuming that the $D$ meson is heavy enough for duality to hold between the
partonic rate and the sum over hadronic final states, one explicitly examines
certain exclusive decays.  This is particularly interesting for studying
$\Delta\Gamma$, which depends on real final states. However, $D$ decays are not
dominated by a small number of final states. Since there are cancellations
between states within a given $SU(3)$ multiplet, one needs to know the
contribution of each state with high precision.  In the absence of sufficiently
precise data on many rates and on strong phases, one is forced to use some
assumptions.  While most studies find $x,y \lesssim 10^{-3}$, it has also been
argued that $SU(3)$ violation is of order unity and so $x,y \sim 10^{-2}$ is
possible.\,\cite{wolf,cnp,kaeding} 

The importance of $SU(3)$ cancellations in both the magnitudes and phases of
matrix elements is nicely illustrated by two-body $D$ decays to charged
pseudoscalars ($\pi^+\pi^-$, $\pi^+ K^-$, $K^+ \pi^-$, $K^+ K^-$).  It is known
that $SU(3)$  breaking is sizable in certain decay rates, e.g., $\BR(D^0 \to
K^+ K^-) / \BR(D^0 \to \pi^+ \pi^-) \simeq 2.8$,\,\cite{PDG} whereas it should
be unity in the $SU(3)$ limit.  Such effects were the basis for the claim that
$SU(3)$ is not applicable to $D$ decays.\,\cite{wolf}${}^,$\,\footnote{The
$SU(3)$ breaking in the matrix elements may be modest, even though the ratio of
the measured rates appears to be very far from the $SU(3)$
limit.\,\cite{MJS}}\,  In contrast, we know very little about the strong phase
$\delta$ between the Cabibbo allowed and doubly Cabibbo suppressed amplitudes,
which vanishes in the $SU(3)$ limit.  Some model calculation give $\cos\delta
\gtrsim 0.8$,\,\cite{Falk:1999ts} but it is possible to obtain much larger
results for $\delta$.\,\cite{cnp}\,  The value of $y$ corresponding to two-body
decays to charged $\pi$'s and $K$'s is
\beq
y_{\pi K} = \BR(D^0 \to \pi^+\pi^-) + \BR(D^0 \to K^+K^-) -
  2\cos\delta\, \sqrt{\BR(D^0 \to K^-\pi^+)\, \BR(D^0 \to K^+\pi^-)} \,.
\eeq
The experimental central values, allowing for $D$ mixing in the doubly Cabibbo
suppressed rates, yield $y_{\pi K} \simeq (5.76 - 5.29 \cos\delta) \times
10^{-3}$.\,\cite{BGLNP}\,  For small $\delta$, there is an almost perfect
cancellation, even though the individual rates violate $SU(3)$ significantly. 
In this ``exclusive'' approach, $x$ is usually obtained from $y$ using a
dispersion relation, and one typically finds $x\sim y$.

At the present time, one cannot use the exclusive approach to reliably predict
$x$ or $y$, since the estimates depend very sensitively on $SU(3)$ breaking in
poorly known strong phases and doubly Cabibbo suppressed rates.  While
calculating these effects model independently is not tractable in general, one
source of $SU(3)$ breaking in $y$, from final state phase space, can be
calculated with only minimal assumptions.  We estimate these effects in the
next section.

\section{$SU(3)$ breaking from phase space}

There is a contribution to the $D^0$ width difference from all final states
common in $D^0$ and $\D0bar$ decay.  In the $SU(3)$ limit these contributions
cancel when one sums over complete $SU(3)$ representations.  The cancellations
depend on $SU(3)$ symmetry both in the matrix elements and in the final state
phase space.  Since model independent calculations of $SU(3)$ violation in
matrix elements are not available, we focus on $SU(3)$ violation in the phase
space.  This depends only on the hadron masses in the final state, and can be
computed with mild assumptions about the momentum dependence of the matrix
elements.  Below we estimate $y$ solely from $SU(3)$ violation in the phase
space.\,\footnote{The phase space difference alone can explain the large
$SU(3)$ breaking between the measured $D\to K^*\ell\bar\nu$ and $D\to
\rho\ell\bar\nu$ rates, assuming no $SU(3)$ breaking in the form
factors.\,\cite{LSW}\,  Recently it was shown that the lifetime ratio of the
$D_s$ and $D^0$ mesons may also be explained this way.\,\cite{NP}}  We find
that this contribution to $y$ is negligible for two-body pseudoscalar final
states, but can be of the order of a percent for final states with mass
near~$m_D$.

Let us concentrate on final states $F$ which transform in a single $SU(3)$
representation $R$.  Assuming $CP$ symmetry in $D$ decays, which in the
Standard Model and in most new physics scenarios is an excellent approximation,
relates $\langle \D0bar| {\cal H}_w |n\rangle$ to $\langle D^0| {\cal H}_w
|\overline{n} \rangle$. Since $|n\rangle$ and $|\overline{n}\rangle$ are in the
same $SU(3)$ multiplet, these two matrix elements are determined by a single
effective Hamiltonian.  Hence the contribution of the states $n\in F_R$ to $y$
is
\beq\label{yfr}
y_{F,R} = {1\over\Gamma}\, \langle\D0bar|\,{\cal H}_w
  \bigg\{ \etacp(F_R)\sum_{n \in F_R}
  |n\rangle \rho_n\langle n| \bigg\} {\cal H}_w\,|D^0\rangle
= {\sum_{n\in F_R} \langle\D0bar|\,{\cal H}_w|n\rangle \rho_n
  \langle n|{\cal H}_w\,|D^0\rangle \over \sum_{n\in F_R}
  \Gamma(D^0\to n)}\,,
\eeq
where $\rho_n$ is the phase space available to the state $n$, and $\etacp = \pm
1$ is determined by the $CP$ transformation of the final state, $CP|n\rangle =
\etacp|\bar n\rangle$.  In the $SU(3)$ limit, the
$\rho_n$'s are the same for $n\in F_R$.  Since $\rho_n$ depend only on the
known masses of the particles in the state $n$, incorporating the true values
of $\rho_n$ in the sum is a calculable source of $SU(3)$ breaking.

\begin{table}
\caption{Values of $y_{F,R}$ for two-body final states.  This represents
the value which $y$ would take if elements of $F_R$ were the only
channel open for $D^0$ decay.}
\label{ytwobody}
\vspace{4pt}
\begin{center}
\begin{tabular}{|@{~~~}lccc|} \hline
\multicolumn{2}{|c}{Final state representation}
  &  $y_{F,R}/s_1^2$  &  $y_{F,R}\ (\%)$  \\ \hline
$PP$  &  $8$  &  $-0.0038$ & $-0.018$  \\
  &  $27$  &  $-0.00071$  & $-0.0034$ \\ \hline
$PV$  &  $8_S$  &  $0.031$ & $0.15$\\
  &  $8_A$  &  $0.032$  & $0.15$ \\
  &  $10$  &  $0.020$ & $0.10$ \\
  &  $\overline{10}$  &  $0.016$ & $0.08$ \\
  &  $27$  &  $0.040$  & $0.19$\\ \hline
$(VV)_{\mbox{\small $s$-wave}}$  &  $8$  &  $-0.081$ & $-0.39$ \\
  &  $27$  &  $-0.061$ & $-0.30$\\
$(VV)_{\mbox{\small $p$-wave}}$  &  $8$  &  $-0.10$ & $-0.48$\\
  &  $27$  &  $-0.14$ & $-0.70$ \\
$(VV)_{\mbox{\small $d$-wave}}$  &  $8$  &  $0.51$ & $2.5$ \\
  &  $27$  &  $0.57$  & $2.8$\\ \hline
\end{tabular}
\end{center}
\end{table}

As the simplest example, consider $D$ decays to states $F=PP$ consisting of a
pair of pseudoscalar mesons such as $\pi$, $K$, $\eta$.  We neglect
$\eta-\eta'$ mixing, but checked that this has a negligible effect on our
numerical results.  Since $PP$ is symmetric in the two mesons, it must
transform as an element of $(8\times 8)_S=27+8+1$.  It is straightforward to
construct the $SU(3)$ invariants, and compute $y_{PP,R}$ from them.  For
example, for the $PP$ in an 8, there are invariants with ${\cal H}_w$ in a
$\15bar$, $A_8^{\15bar}(PP_8)^k_iH^{ij}_kD_j$, and with ${\cal H}_w$ in a $6$,
$A_8^6(PP_8)^k_iH^{ij}_kD_j$. In this particular case, the product
$H^{ij}_kD_j$ with $(ij)$ symmetric (the $\15bar$) is proportional to
$H^{ij}_kD_j$ with $(ij)$ antisymmetric (the 6), and the linear combination
$A_8\equiv A_8^{\15bar}-A_8^6$ is the only one which appears.  We find
\beqa\label{ypp8}
&&{}\!\!\! y_{PP,8} \propto s_1^2\,\bigg[ \frac12\, \Phi(\eta,\eta)
  + \frac12\,  \Phi(\pi^0,\pi^0) + \frac13\, \Phi(\eta,\pi^0)
  + \Phi(\pi^+,\pi^-) + \Phi(K^+,K^-)  \\*
&&{}\!\!\! - \frac16\,\Phi(\eta,K^0) - \frac16\, \Phi(\eta,\K0bar) 
  - \Phi(K^+,\pi^-) - \Phi(K^-,\pi^+) - \frac12\, \Phi(K^0,\pi^0)
  - \frac12\, \Phi(\K0bar,\pi^0) \bigg] ,\nonumber
\eeqa
where $\Phi(P_1,P_2)$ is the phase space for $D^0\to P_1 P_2$ decay.  In the
$SU(3)$ limit all $\Phi$'s are equal, and $y_{PP,8}$ vanishes as $m_s^2$.
In a two-body decay $\Phi(P_1,P_2)$ is proportional to $|\vec p\,|^{2\ell+1}$,
where $\vec p$ and $\ell$ are the spatial momentum and orbital angular momentum
of the final state particles.  For $D^0\to PP$, the decay is into an $s$-wave,
and it is straightforward to compute the phase space factors from the
pseudoscalar masses.  The results are not larger than one finds in the
inclusive analysis (see Table~\ref{ytwobody}), since as in the parton picture,
the final states are far from threshold.

Next we turn to final states consisting of a pseudoscalar and a vector meson,
$F=PV$.  In this case there is no symmetry between the mesons, so all
representations in the combination $8\times 8 = 27 +10 +\tenbar +8_S +8_A +1$
can appear.  For simplicity, we take the quark content of the $\phi$ and
$\omega$ to be $\bar ss$ and $(\bar uu+\bar dd)/\sqrt2$ respectively, and
consider only the combination which appears in the $SU(3)$ octet.  Reasonable
variations of the $\phi - \omega$ mixing angle have a negligible effect on our
results.  Both because the vector mesons are more massive, and because the
decay is now into a $p$-wave, the phase space effects are larger than for the
$PP$ final state (see Table~\ref{ytwobody}). Still, for all representation,
$y_{PV}$ are less than a percent.  Note that three-body final states $3P$ can
resonate through $PV$, and so are partially included here. 

For the $VV$ final state, decays into $s$-, $p$- and $d$-waves are all
possible. Bose symmetry and the restriction to zero total angular momentum
together imply that only the symmetric $SU(3)$ combinations appear.  Because
some $VV$ final states, such as $\phi K^*$, lie near the $D$ threshold, the
vector meson widths are very important.  We model the resonance line shapes
with Lorentz invariant Breit-Wigner distributions, $m^2\,\Gamma_R^2 /
[(m^2-m_R^2)^2 + m^2\,\Gamma_R^2]$, where $m_R$ and $\Gamma_R$ are the mass and
width of the vector meson, and $m^2$ is the square of its four-momentum in the
decay. The results for $s$-, $p$-, and $d$-wave decays are shown in
Table~\ref{ytwobody}. With these heavier final states and with the higher
partial waves, effects at the level of a percent are quite generic.  If the
vector meson widths were neglected, the results in the $p$- and $d$-wave
channels would be larger by approximately a factor of three.  The finite widths
soften the $SU(3)$ breaking which would otherwise be induced by a sharp phase
space boundary.  Again, $4P$ and $PPV$ final states can resonate through $VV$,
so they are partially included here.

As we go to final states with more particles, the combinatoric possibilities
begin to proliferate.  We will only consider the final states $3P$ and $4P$,
and require that the pseudoscalars are in a totally symmetric 8 or 27
representation of $SU(3)$. This assumption is convenient, because the phase
space integration is much simpler if it can be performed symmetrically.  We
have no reason to believe that this choice selects final state multiplets for
which phase space effects are particularly enhanced or suppressed.  The results
for $y_{3P}$ and $y_{4P}$ are shown in Table~\ref{mltybodytbl}.

\begin{table}
\caption{Values of $y_{F,R}$ for some three- and four-body final states.}
\label{mltybodytbl}
\vspace{4pt}
\begin{center}
\begin{tabular}{|@{~~~}lccc|} \hline
\multicolumn{2}{|c}{Final state representation} 
  &  $y_{F,R}\,/s_1^2$  &  $y_{F,R}\ (\%)$ \\ \hline
$(3P)_{\mbox{\small $s$-wave}}$	&  $8$  &  $-0.48$  & $-2.3$\\
  &  $27$  &  $-0.11$  & $-0.54$ \\
$(3P)_{\mbox{\small $p$-wave}}$	&  $8$  &  $-1.13$  & $-5.5$ \\
  &  $27$  &  $-0.07$  & $-0.36$  \\
$(3P)_{\mbox{\small form-factor}}$	&  $8$  &  $-0.44$  & $-2.1$\\
  &  $27$  &  $-0.13$ & $-0.64$ \\ \hline
$4P$  &  $8$  &  $3.3$ & $16$  \\
  &  $27$  &  $2.2$  & $11$ \\
  &  $27'$  &  $1.9$ & $9.2$ \\ \hline
\end{tabular} 
\end{center}
\end{table}

In contrast to the two-body decays, for three-body final states the momentum
dependence of the matrix elements is no longer fixed by angular momentum
conservation.  The simplest assumption is to take a momentum independent matrix
element, with all three final state particles in an $s$-wave.  We have also
considered other matrix elements; for example, if one of the mesons has angular
momentum $\ell = 1$ in the $D^0$ rest frame (balanced by the combination of the
other two mesons).  One could also imagine introducing a mild ``form factor
suppression,'' with a weight such as $\Pi_{i\ne j}(1 - m_{ij}^2 / Q^2)^{-1}$,
where $m_{ij}^2 = (p_i+p_j)^2$, and $Q = 2\,$GeV is a typical resonance mass. 
The resulting $y_{F,R}$ values are show in Table~\ref{mltybodytbl}.

Finally, we studied $4P$ final states with the mesons in fully symmetric 8 or
27.  The results for momentum independent matrix elements are summarized in
Table~\ref{mltybodytbl}.  (Note that the last two entries in the last column
were mixed up in our paper.\,\cite{FGLP}).  There are actually two symmetric
27's; we call 27 and $27'$ the representations of the form $R^{ij}_{kl} =
[M^i_mM^m_kM^j_nM^n_l+ {\rm symmetric} - {\rm traces}]$ and $R^{ij}_{kl} =
[M^i_mM^m_nM^n_kM^j_l + {\rm symmetric} - {\rm traces}]$, respectively.  Here
the partial contributions to $y$ are very large, of the order of 10\%. This is
not surprising, since $4P$ final states containing more than one strange
particle are close to $D$ threshold, and the ones with no pions are
kinematically inaccessible.  So there is no reason for $SU(3)$ cancellations to
persist effectively.

Formally, one can construct $y$ from the individual $y_{F,R}$ by weighting
them by their $D^0$ branching ratios,
\beq\label{ycombine}
y = {1\over\Gamma} \sum_{F,R}\, y_{F,R}
  \bigg[\sum_{n\in F_R} \Gamma(D^0\to n)\bigg]\,.
\eeq
However, the data on $D$ decays are neither abundant nor precise enough to
disentangle the decays to the various $SU(3)$ multiplets, especially for the
three- and four-body final states.  Nor have we computed $y_{F,R}$ for all or
even most of the available representations.  Thus, we can only estimate $y$ by
assuming that the representations for which we know $y_{F,R}$ are typical for
final states with a given multiplicity, and scale to the branching ratio of
those final states.  Table~\ref{branching} summarizes the $D^0$ branching
ratios to two-, three- and four-body final states, rounded to the nearest 5\%,
to emphasize the uncertainties in the data.\,\cite{PDG}\,  Almost half of the
$D^0$ width is accounted for.  Based on data in the ${\overline
K}{}^{0*}\rho^0$ channel, the $VV$ final state is dominantly $CP$ even,
consistent with an equal distribution of $s$- and $d$-wave decays.  Taking the
product of the typical $y_{F,R}$ with the approximate branching ratios in
Table~\ref{branching}, we can estimate the contribution to $y$ from a given
type of final state.  While in most cases the contributions are small, of the
order of $10^{-3}$ or less, $D^0$ decays to nonresonant $4P$ states naturally
contribute to $y$ at the percent level.  The reason for so large $SU(3)$
violating effects in $y$ is that a sizable fraction of $D^0$ decays are to
final states for which the complete $SU(3)$ multiplets are not kinematically
accessible.

\begin{table}
\caption{Total $D^0$ branching fractions to classes of final states,
rounded to the nearest 5\%.\,\protect\cite{PDG}}
\label{branching}
\vspace{4pt}
\begin{center}
\begin{tabular}{|ccccccc|} \hline
Final state  &  $PP$ & $PV$ & $(VV)_{\mbox{\small $s$-wave}}$
  & $(VV)_{\mbox{\small $d$-wave}}$ & $3P$ & $4P$  \\ \hline
Branching fraction & 5\% & 10\% & 5\% & 5\% & 5\% & 10\% \\ \hline
\end{tabular}
\end{center}
\end{table}

We have not considered all possible final states which may give large 
contributions to $y$.  For example, $\BR(D^0\to  K^-a_1^+) =
(7.3\pm1.1)\%$,\,\cite{PDG} although the available phase space is very small. 
Unfortunately, the identities of the $SU(3)$ partners of the $a_1(1260)$, which
has $J^{PC}=1^{++}$, are not well established.  While it is natural to identify
the $K_1(1400)$ and $f_1(1285)$ as the analogues of the $K^*$ and $\omega$,
respectively, there is no natural candidate for the $\bar ss$ analogue of the
$\phi$.  The value of $y_{PV^*}$ is also sensitive to the poorly known width of
the $a_1$.  If we take the $\bar ss$ state to be the $f_1(1420)$, and
$\Gamma(a_1) = 400\,$MeV, we find $y_{PV^*,8_S} = 1.8\%$.  If instead we take
the $f_1(1510)$, we find $y_{PV^*,8_S} = 1.7\%$.  With $\Gamma(a_1) =
250\,$MeV, these numbers become $2.5\%$ and $2.4\%$, respectively.  Although
percent level contributions to $y$ are clearly possible from this channel, the
data are still too poor to draw firm conclusions.

From our analysis, in particular as applied to the $4P$ final state, we
conclude that $y$ of the order of a percent appears completely natural.  An
order of magnitude smaller result would require significant cancellations which
would only be expected if they were enforced by the OPE.  The hypothesis
underlying the present analysis is that this is not the case.

\section{Conclusions}

The motivation most often cited in searches for $D^0 - \D0bar$ mixing is the
possibility of observing a signal from new physics which may dominate over the
Standard Model contribution.  But to look for new physics in this way, one must
be confident that the Standard Model prediction does not already saturate the
experimental bound.  Previous analyses based on short distance expansions have
consistently found $x,y\sim10^{-3}$, while naive estimates based on known
$SU(3)$ breaking in charm decays allow an effect an order of magnitude larger.
Since current experimental sensitivity is at the level of a few percent, the
difference is quite important.

We proved that if $SU(3)$ violation can be treated perturbatively, then $D^0 -
\D0bar$ mixing in the Standard Model is generated only at second order in
$SU(3)$ breaking effects. Within the exclusive approach, we identified an
$SU(3)$ breaking effect, $SU(3)$ violation in final state phase space, whose
contribution to $y$ can be calculated with small model dependence.  We found
that phase space effects in $D$ decays to final states near threshold can
induce $y \sim 10^{-2}$.

The implication of our results for the Standard Model prediction for $x$ is
less apparent. While analyses based on the ``inclusive" approach generally
yield $x \gtrsim y$, it is not clear what the ``exclusive" approach predicts.
If $x > y$ is found experimentally, it may still be an indication of a new
physics contribution to $x$, even if $y$ is also large.  On the other hand, if
$y>x$ then it will be hard to find signals of new physics, even if such
contributions dominate $\Delta M$.  The linear sensitivity to new physics in
the analysis of the time dependence of $D^0\to K^+\pi^-$ is from $x' = x
\cos\delta + y \sin\delta$ and $y'= y \cos\delta - x \sin\delta$ instead of $x$
and $y$.  If $y> x$, then $\delta$ would have to be known precisely for these
terms to be sensitive to new physics in $x$.

There remain large uncertainties in the Standard Model predictions of $x$ and
$y$, and values near the current experimental bounds cannot be ruled out. 
Therefore, it will be difficult to find a clear indication of physics beyond
the Standard Model in $D^0 - \D0bar$ mixing measurements.  We believe that at
this stage the only robust potential signal of new physics in $D^0 - \D0bar$
mixing is $CP$ violation, for which the Standard Model prediction is very
small.  Unfortunately, if $y > x$, then the observable $CP$ violation in $D^0 -
\D0bar$ mixing is necessarily small, even if new physics dominates $x$.  Thus,
to disentangle new physics from Standard Model contributions, it will be
crucial to (i) improve the measurements of both $x$ and $y$; (ii) extract the
relevant strong phase in the time dependence of doubly Cabibbo suppressed
decays; and (iii) look for $CP$ violation, which remains a potentially robust
signal of new physics.

\section*{Acknowledgments}
It is a pleasure to thank Sven Bergmann, Adam Falk, Yuval Grossman, Yossi Nir,
and Alexey Petrov for very enjoyable collaborations on the topics discussed in
this talk.  Special thanks to Alain Blondel for much fun (and some
intimidation...) off-piste.
This work was supported in part by the Director, Office of Science, Office of
High Energy and Nuclear Physics, Division of High Energy Physics, of the U.S.\
Department of Energy under Contract DE-AC03-76SF00098.


\section*{References}
\begin{thebibliography}{99}


\bibitem{FGLP}
A.F.~Falk, Y.~Grossman, Z.~Ligeti and A.A.~Petrov,
Phys.\ Rev.\ D65, 054034 (2002).

\bibitem{BGLNP}
S.~Bergmann, Y.~Grossman, Z.~Ligeti, Y.~Nir, A.A.~Petrov,
Phys.\ Lett.\ B486, 418 (2000).

\bibitem{sqa}
Y.~Nir and N.~Seiberg, Phys.\ Lett.\ B309, 337 (1993);
Z.~Ligeti and Y.~Nir, hep-ph/0202117.

\bibitem{Ge92}
H. Georgi, Phys.\ Lett.\ B297, 353 (1992).

\bibitem{Oh93}
T. Ohl, G. Ricciardi and E.H. Simmons, Nucl.\ Phys.\ B403, 605 (1993).

\bibitem{Da85}
A. Datta and M. Khambakhar, Zeit.\ Phys.\ C27, 515 (1985).

\bibitem{dght}
J. Donoghue, E. Golowich, B. Holstein and J. Trampetic,
Phys.\ Rev.\ D33, 179 (1986).

\bibitem{e791y}
E.M. Aitala {\it et al.}, E791 Collaboration, Phys.\ Rev.\ Lett.\ 83, 32
(1999).

\bibitem{focusy}
J.M. Link {\it et al.}, FOCUS Collaboration, Phys.\ Lett.\ B485, 62 (2000).

\bibitem{cleoy}
D. Cronin-Hennessy {\it et al.}, CLEO Collaboration, hep-ex/0102006.

\bibitem{belley}
K. Abe {\it et al.}, BELLE Collaboration, Phys.\ Rev.\ Lett.\ 88, 162001
(2002).

\bibitem{babary}
A. Pompili, BABAR Collaboration, these proceedings, hep-ex/0205071.

\bibitem{Godang:2000yd}
R.~Godang {\it et al.}, CLEO Collaboration,
Phys.\ Rev.\ Lett.\ 84, 5038 (2000).

\bibitem{brand}
G.~Brandenburg {\it et al.}, CLEO Collaboration,
Phys.\ Rev.\ Lett.\ 87, 071802 (2001).

\bibitem{DmixSL}
E.M.~Aitala {\it et al.}, E791 Collaboration, Phys.\ Rev.\ D57, 13 (1998).

\bibitem{nelson}
H.N. Nelson, hep-ex/9908021.

\bibitem{Bigi:2000wn}
I. Bigi and N. Uraltsev, Nucl.\ Phys.\ B592, 92 (2000).

\bibitem{wolf}
L. Wolfenstein, Phys.\ Lett.\ B164, 170 (1985).

\bibitem{cnp}
P. Colangelo, G. Nardulli and N. Paver, Phys.\ Lett.\ B242, 71 (1990).

\bibitem{kaeding}
T.A. Kaeding, Phys.\ Lett.\ B357, 151 (1995).

\bibitem{PDG}
D.E.~Groom {\it et al.}, Particle Data Group, Eur.\ Phys.\ J.\ C15, 1 (2000).

\bibitem{MJS}
M.J. Savage, Phys.\ Lett.\ B257, 414 (1991).

\bibitem{Falk:1999ts}
A.F.~Falk, Y.~Nir and A.A.~Petrov, JHEP 12, 019 (1999).

\bibitem{LSW}
Z.~Ligeti, I.W.~Stewart and M.B.~Wise, Phys.\ Lett.\ B420, 359 (1998).

\bibitem{NP}
S. Nussinov and M.V. Purohit, Phys.\ Rev.\ D65, 034018 (2002).

\end{thebibliography}
\end{document}